\documentclass[twocolumn]{aastex62}

\usepackage{amsmath,amsbsy}
\usepackage{hyperref}

\received{\today}

\shorttitle{Energetic Particles and Coherent Structures}
\shortauthors{Bandyopadhyay et al.}

\newcommand{\isois}{IS$\odot$IS }

\begin{document}
	
\title{Observations of Energetic-Particle Population Enhancements along Intermittent Structures near the Sun from Parker Solar Probe}

\author[0000-0002-6962-0959]{Riddhi Bandyopadhyay}
\affiliation{Department of Physics and Astronomy, University of Delaware, Newark, DE 19716, USA}

\author[0000-0001-7224-6024]{W.~H. Matthaeus}
\email{whm@udel.edu}
\affiliation{Department of Physics and Astronomy, University of Delaware, Newark, DE 19716, USA}
\affiliation{Bartol Research Institute, University of Delaware, Newark, DE 19716, USA}

\author[0000-0003-0602-8381]{T.~N. Parashar}
\affiliation{Department of Physics and Astronomy, University of Delaware, Newark, DE 19716, USA}
\affiliation{Bartol Research Institute, University of Delaware, Newark, DE 19716, USA}

\author[0000-0002-7174-6948]{R. Chhiber}
\affiliation{Department of Physics and Astronomy, University of Delaware, Newark, DE 19716, USA}
\affiliation{NASA Goddard Space Flight Center, Greenbelt, MD 20771, USA}

\author[0000-0003-3414-9666]{D. Ruffolo}
\affiliation{Department of Physics, Faculty of Science, Mahidol University, Bangkok 10400, Thailand}

\author[0000-0002-5317-988X]{M.~L. Goldstein}
\affiliation{NASA Goddard Space Flight Center, Greenbelt, MD 20771, USA}
\affiliation{University of Maryland Baltimore County, Baltimore, MD 21250, USA}

\author[0000-0002-2229-5618]{B.~A. Maruca}
\affiliation{Department of Physics and Astronomy, University of Delaware, Newark, DE 19716, USA}
\affiliation{Bartol Research Institute, University of Delaware, Newark, DE 19716, USA}
		
\author[0000-0001-8478-5797]{A. Chasapis}
\affiliation{Laboratory for Atmospheric and Space Physics, University of Colorado Boulder, Boulder, CO 80303, USA}

\author[0000-0001-8358-0482]{R. Qudsi}
\affiliation{Department of Physics and Astronomy, University of Delaware, Newark, DE 19716, USA}

\author{D.~J. McComas} 
\affiliation{Department of Astrophysical Sciences, Princeton University, Princeton, NJ, 08544, USA}

\author{E.~R. Christian}
\affiliation{NASA Goddard Space Flight Center, Greenbelt, MD 20771, USA}

\author[0000-0003-2685-9801]{J.~R. Szalay}
\affiliation{Department of Astrophysical Sciences, Princeton University, Princeton, NJ, 08544, USA}

\author{C.~J. Joyce}
\affiliation{Department of Astrophysical Sciences, Princeton University, Princeton, NJ, 08544, USA}

\author{J. Giacalone}
\affiliation{University of Arizona, Tucson, AZ 85721, USA}

\author{N.~A. Schwadron}
\affiliation{University of New Hampshire, Durham, NH, 03824, USA}
\affiliation{Department of Astrophysical Sciences, Princeton University, Princeton, NJ, 08544, USA}

\author{D.~G. Mitchell}
\affiliation{Johns Hopkins University Applied Physics Laboratory, Laurel, MD 20723, USA}

\author{M.~E. Hill}
\affiliation{Johns Hopkins University Applied Physics Laboratory, Laurel, MD 20723, USA}

\author{M.~E. Wiedenbeck}
\affiliation{California Institute of Technology, Pasadena, CA 91125, USA}

\author{R.~L. McNutt Jr.}
\affiliation{Johns Hopkins University Applied Physics Laboratory, Laurel, MD 20723, USA}

\author{M. I. Desai}
\affiliation{University of Texas at San Antonio, San Antonio, TX 78249, USA}

\author[0000-0002-1989-3596]{Stuart D. Bale}
\affil{Physics Department, University of California, Berkeley, CA 94720-7300, USA}
\affil{Space Sciences Laboratory, University of California, Berkeley, CA 94720-7450, USA}
\affil{The Blackett Laboratory, Imperial College London, London, SW7 2AZ, UK}
\affil{School of Physics and Astronomy, Queen Mary University of London, London E1 4NS, UK}

\author[0000-0002-0675-7907]{J. W. Bonnell}
\affil{Space Sciences Laboratory, University of California, Berkeley, CA 94720-7450, USA}

\author[0000-0002-4401-0943]{Thierry {Dudok de Wit}}
\affil{LPC2E, CNRS and University of Orl\'eans, Orl\'eans, France}

\author[0000-0003-0420-3633]{Keith Goetz}
\affiliation{School of Physics and Astronomy, University of Minnesota, Minneapolis, MN 55455, USA}

\author[0000-0002-6938-0166]{Peter R. Harvey}
\affil{Space Sciences Laboratory, University of California, Berkeley, CA 94720-7450, USA}

\author[0000-0003-3112-4201]{Robert J. MacDowall}
\affil{Solar System Exploration Division, NASA/Goddard Space Flight Center, Greenbelt, MD, 20771}

\author[0000-0003-1191-1558]{David M. Malaspina}
\affil{Laboratory for Atmospheric and Space Physics, University of Colorado, Boulder, CO 80303, USA}

\author[0000-0002-1573-7457]{Marc Pulupa}
\affil{Space Sciences Laboratory, University of California, Berkeley, CA 94720-7450, USA}

\author{M. Velli}
\affiliation{Department of Earth, Planetary, and Space Sciences, University of California, Los Angeles, CA 90095, USA}

\author[0000-0002-7077-930X]{J.C. Kasper}
\affiliation{Climate and Space Sciences and Engineering, University of Michigan, Ann Arbor, MI 48109, USA}
\affiliation{Smithsonian Astrophysical Observatory, Cambridge, MA 02138 USA.}

\author[0000-0001-6095-2490]{K.E. Korreck}
\affiliation{Smithsonian Astrophysical Observatory, Cambridge, MA 02138 USA.}

\author[0000-0002-7728-0085]{M. Stevens}
\affiliation{Smithsonian Astrophysical Observatory, Cambridge, MA 02138 USA.}

\author[0000-0002-3520-4041]{A.W. Case}
\affiliation{Smithsonian Astrophysical Observatory, Cambridge, MA 02138 USA.}

\author{N. Raouafi}
\affiliation{Johns Hopkins University Applied Physics Laboratory, Laurel, MD 20723, USA}

	
	
\begin{abstract}
Observations at 1 au have confirmed that enhancements in measured energetic particle fluxes are statistically associated with ``rough'' magnetic 
fields, i.e., fields having atypically large spatial derivatives or increments,
as measured by the Partial Variance of Increments (PVI) method.
One way to interpret this observation is as an association of the energetic particles with trapping or channeling 
within magnetic flux tubes, possibly near their boundaries. 
However, it remains unclear whether this association is a transport or local effect;
i.e., the particles might have been energized at a distant location, perhaps by shocks or reconnection,
or they might experience local energization or re-acceleration.  
The Parker Solar Probe (PSP), even in its first two orbits, offers a unique opportunity to 
study this statistical correlation closer to the corona. As a first step, we analyze the separate correlation properties of the energetic particles measured by the \isois instruments during the first solar encounter. The distribution of time intervals between a specific type of event, 
i.e., the waiting time, can indicate the nature of the underlying process.
We find that the \isois observations show a power-law distribution of waiting times, indicating a correlated (non-Poisson) distribution.  Analysis of low-energy \isois data suggests that the results are consistent with the 1 au studies, although we find hints of some unexpected behavior. A more complete understanding of these statistical distributions will provide valuable insights into the origin and propagation of solar energetic particles, a picture that should become clear with future PSP orbits.
\end{abstract}



\section{Introduction} \label{sec:intro}
The transport and acceleration of charged energetic particles (EP) is well known to be intimately related to the properties of plasma turbulence~\citep{Jokipii66}. Transport is particularly sensitive to the magnetic field  
structure~\citep{Seripienlert2010ApJ,Tooprakai2016ApJ,Malandraki2019ApJ}, as well as the distribution of fluctuations over 
scale~\citep{BieberEA94}. Propagation of particles gives 
rise to a complex relationship of the particle trajectory with the electric fields that ultimately 
accounts for acceleration in space and astrophysics \citep{Terasawa1989Science,Reames1999SSR,Amato2018ASR}. 
While many features of energetic particles
can be understood in theoretical frameworks based
on quasi-linear theory and random-phase fluctuations, 
there is an increasing interest in
phenomena that can only be understood by taking into account 
coherent magnetic structures.
Here we mean the organization of turbulent magnetic fields into flux tubes, and their associated coherent structures, such as current sheets, current cores, and secondary flux tubes, plasmoids, and islands that are frequently found on, or near, borders between interacting flux tubes. Moreover,
the turbulence cascade can give rise to magnetic 
``islands'' or secondary flux tubes, 
having dimensions that span a wide range of length scales \citep{WanEA13-xpoints, ZhdankinEA12, ZhdankinEA13, LoureiroEA12}.
Such structures have been 
suggested to have major effects on charged
particle populations, including transport,
energization, or both~\citep{Khabarova2015ApJ, Khabarova2017ApJ}. 
Most of these physical effects and theoretical constructs 
have found application in the description of Solar Energetic Particles in the heliosphere, and it is fair to say that many questions remain incompletely settled. One reason is that the different mechanisms and effects cannot always be distinguished at 1 au and beyond, 
due to the significant 
ambiguity introduced by the intervening transport and solar wind dynamics.

A main goal of the recently launched 
Parker Solar probe Mission (PSP)~\citep{Fox2016SSR} has been to disentangle 
the effects of transport and local acceleration on heliospheric 
energetic particle populations by making observations that lie much closer to the sources or the energetic particles than any prior mission.
Here, we analyze PSP observations from its first two solar encounters to provide a first look at the statistics of energetic 
 particles and their relation to magnetic field roughness or intermittency~\citep{GrecoEA09}. 
 We employ energetic particle data from the \isois instruments
 \citep{McComas2016SSR}, 
 and magnetic field data~\citep{Bale2016SSR} from the FIELDS magnetometers on board PSP.

\begin{figure*}[ht!]
	\begin{center}
		\includegraphics[width=\linewidth]{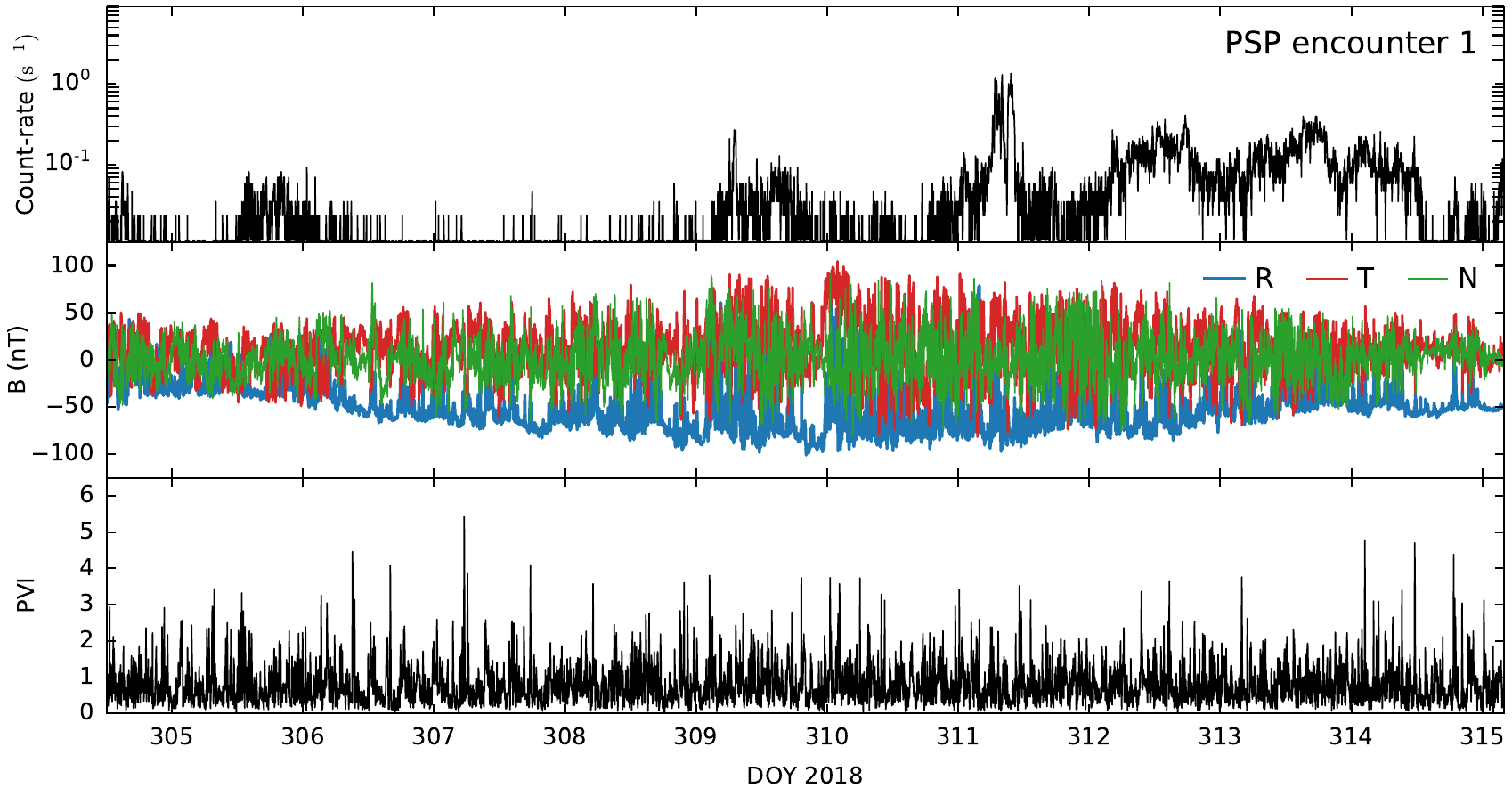}
		\includegraphics[width=\linewidth]{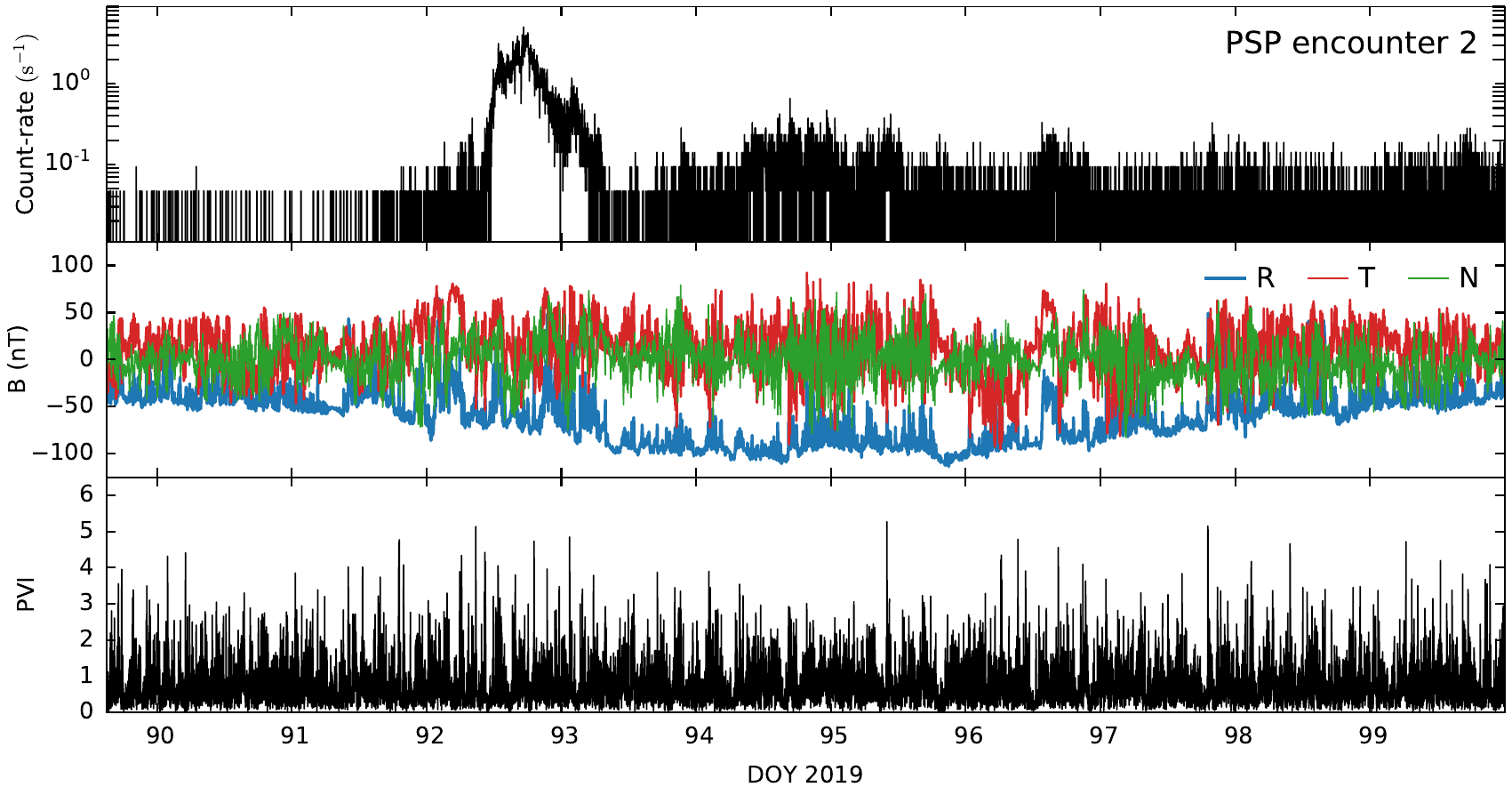}
		\caption{Data from first two encounters of PSP with the Sun: The top panels show the proton count rate, the middle panels plot the RTN components of the magnetic field, and the bottom panels show the PVI.}
		\label{fig:overview}
	\end{center}
\end{figure*}

The essential motivation for the present study comes 
from prior works that found a statistical 
association of measures of particle energization
and measures of 
magnetic discontinuities using the Partial Variance of 
Increments (PVI) method~\citep{GrecoEA09,Greco2017SSR}.
In particular, \cite{OsmanEA11-swHeat} 
found that the solar wind at 1 au is hotter near 
high PVI events (i.e., locally large gradients of the magnetic field). High PVI often corresponds to current structures or current concentrations~\citep{Chasapis2015ApJL, Greco2017SSR}. These may be found in the form of sheets, cores, or other formations and are often seen in simulations at boundaries of interacting flux tubes.
As such, very strong PVI events have been statistically associated with reconnection events in MHD simulation~\citep{Servidio2011JGR} and in the solar wind~\citep{Osman2014PRL}. The PVI method efficiently identifies classical MHD discontinuities, but the method does not distinguish different types of discontinuities such as tangential discontinuities and rotational discontinuities. What we mean by “coherent structure” is simply a concentration of gradients in space. This mathematically requires phase coherence at certain points or regions where the concentration is located. But in principle this idea includes many possible types of structures. Sharp changes in the solar wind magnetic field can be associated with various structures, which may not be turbulence associated, e.g., large-scale current sheets, including the heliospheric current sheet, interplanetary shocks, large-discontinuities associated with solar streamers, Coronal Mass Ejections (CME), and Co-rotating Interaction Regions (CIR) borders, magnetic clouds, magnetic islands inside a fragmented magnetic cloud. However, since we compute the PVI values using a relatively small-scale increment lag (see section~\ref{sec:enc}), it is unlikely that the PVI would be sensitive to such very large scale objects. Subsequently, \citet{Tessein2013ApJL,Tessein2015ApJ} 
found a similar, strong statistical association between high PVI events and enhanced flux of energetic particles using data from the ACE spacecraft. 
More recently, \citet{Khabarova2015ApJ, Khabarova2017ApJ}, and \citet{Malandraki2019ApJ}
found evidence that island-like magnetic structures 
are associated with higher fluxes of energetic particles. 
These observations complement theoretical work ~\citep{AmbrosianoEA88,DmitrukEA04,Zank2014ApJ} 
that describe mechanisms for particle energization 
by trapping in secondary magnetic structures
such as current channels, or small magnetic islands that tend to form during dynamical activity near flux tube boundaries or current sheets. So far, conclusive evidence has not been available to unambiguously distinguish between secondary magnetic structures as transport conduits, a point emphasized by \cite{TesseinEA16}.
Moreover, if secondary islands and flux tubes facilitate and guide transport, there remains the question of where 
and how the transport originates and what the sources of particles are.  For example, is the source close to a nearby CME or interplanetary shock, or is the source  distant, perhaps deep in the corona.   
Although we are not able to answer all these questions in detail,
providing new statistical correlations
relating 
SEPs and PVI events 
during the PSP encounters adds new and important constraints to 
our understanding of the characteristics of heliospheric EP populations and the mechanisms affecting 
them. Our analysis also provides a first look at what will eventually be a much more complete survey at even closer distances to the Sun; a goal that will be achieved during subsequent PSP orbits.

\section{Parker Solar Probe Data}\label{sec:overview}
The 
Parker Solar Probe (PSP) mission \citep{Fox2016SSR}
completed two orbits between launch on 8 August 2018 and 
18 June 2019. The first solar encounter comprised 
approximately ten days on each side of the first perihelion at 35.7 $R_{\odot}$ on 06 November 2018, while the second 
perihelion, also near 35.7 $R_{\odot}$, 
occurred on 4 April 2019.
During this passage, 
the Integrated Science Investigation of the Sun (\isois) instrument suite~\citep{McComas2016SSR} performed detailed 
measurements of solar energetic particles (SEPs)~\citep{McComas:sub}.
The FIELDS instrument made high cadence measurements of the vector magnetic field~\citep{Bale:sub}. 
We analyze 
the energetic particle data 
from the \isois suite, 
particularly EPI-Lo ion count-rate: total ions from $\sim$15-200 keV/nuc with no mass discrimination, but likely dominated by protons, averaged over all 80 look directions
(logical\_source: psp\_isois-epilo\_l2-ic, varname: H\_CountRate\_ChanT). 
The EPI-Lo instrument measures ions and ion composition from $\sim 20\, \mathrm{keV/nucleon} - 15\, \mathrm{MeV}$ total energy.
The magnetic field data 
are analyzed 
at 1 min cadence (dataset: psp-fld-l2-mag-RTN-1min).
We calculate magnetic-field PVI at 2 min lag and 4 hour averaging interval, and resample the calculated PVI time series to the EPI-Lo count-rate. Figure~\ref{fig:overview} shows an overview of the first two encounters.

\section{Energetic Particle Statistics} 
\label{sec:wt}
\begin{figure}
	\begin{center}
		\includegraphics[width=\linewidth]{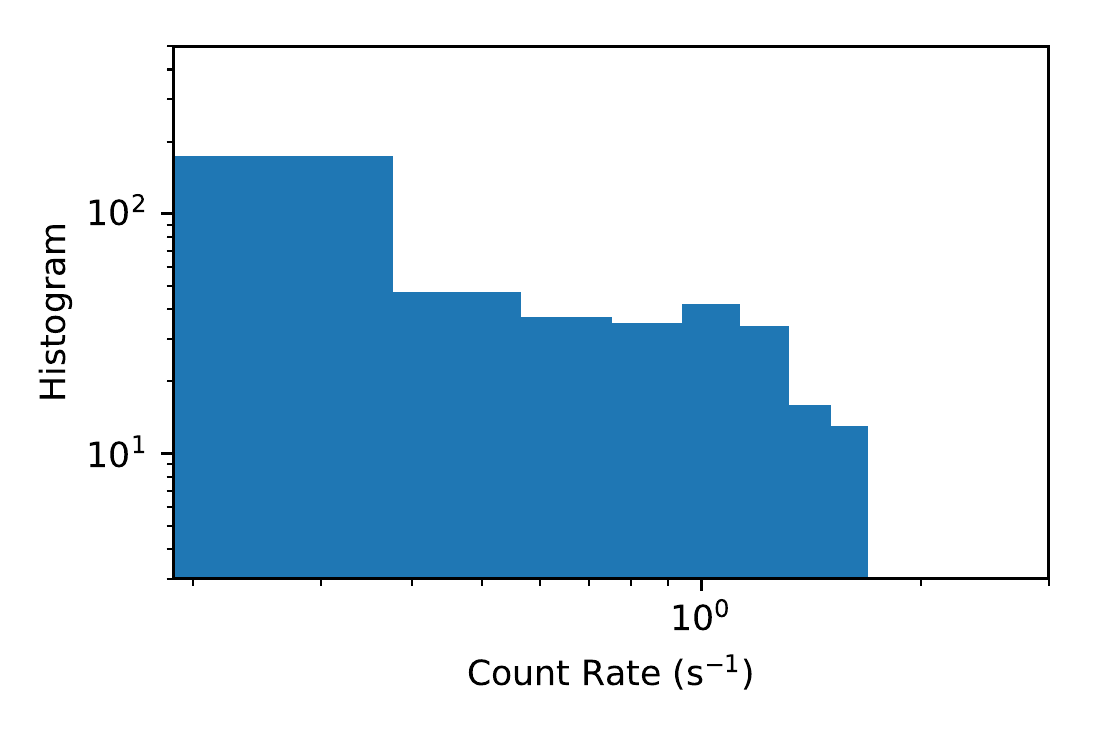}
		\caption{Histograms (showing frequency of occurrence, or number of counts) of count-rates measured by \isois/EPI-Lo for PSP first solar encounter.}
		\label{fig:hist-lo}
	\end{center}
\end{figure}
\begin{figure}
	\begin{center}
		\includegraphics[width=\linewidth]{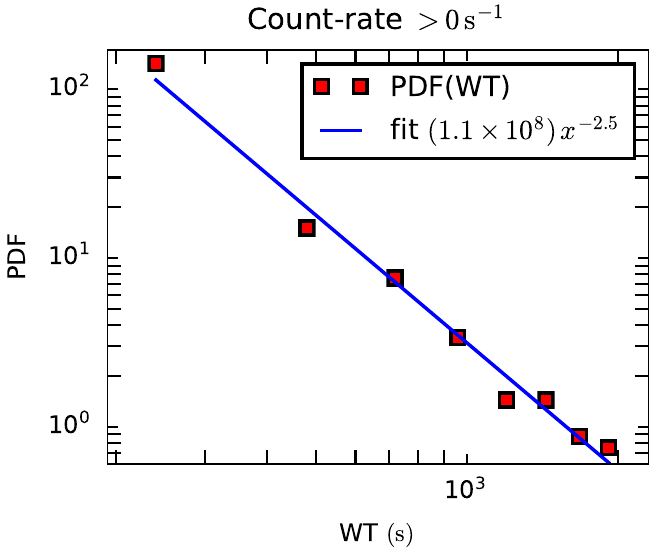}
		\caption{PDF of waiting times between any two non-zero count-rates events for the first encounter. Bins with fewer than five counts are discarded. The power-law fit $a x^{b}$ is shown as a solid, blue line, where $a = (1.1\pm0.9) \times 10^{8}$ and $b = - 2.5 \pm 0.1$.}
		\label{fig:wt0}
	\end{center}
\end{figure}
\begin{figure*}
	\begin{center}
		\includegraphics[width=0.48\linewidth]{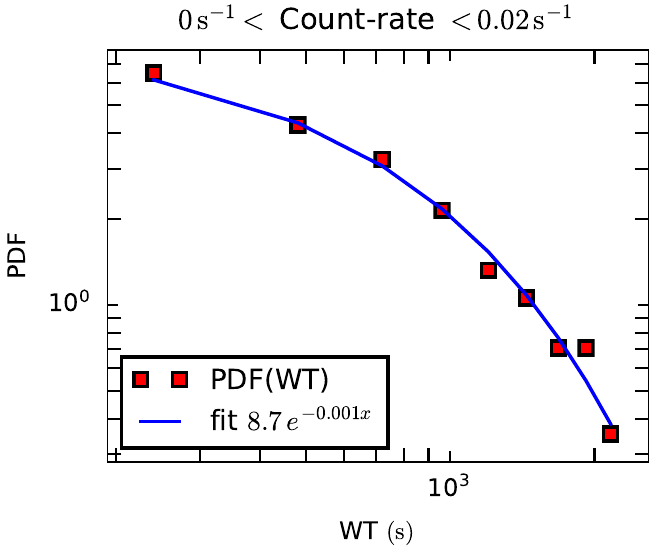}
		\includegraphics[width=0.48\linewidth]{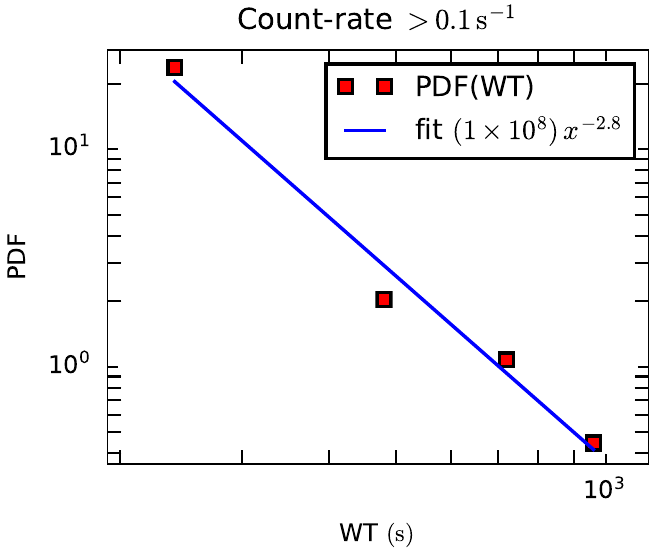}
		\caption{PDFs of waiting times between  events with (left) $0\,\mathrm{s}^{-1} <$ count-rate $< 0.02\,\mathrm{s}^{-1}$ and (right) count-rate $> 0.1\,\mathrm{s}^{-1}$, for the first PSP encounter. The exponential $c e^{-x d}$ and the powerlaw $a x^{b}$ are shown as a solid,blue lines. The parameters and uncertainties of the fit are
		(for the left panel)
		$c = 8.7 \pm 0.8$ and $d = (-144 \pm 7)\times 10^{-5}$; 
		(for the right panel) $a = (1 \pm 1) \times 10^{8}$ and $b = -2.8 \pm 0.3$. Bins with fewer than five counts have been discarded.}
		\label{fig:wt-comb}
	\end{center}
\end{figure*}
Elementary
statistical analysis of random 
signals can reveal a surprising amount of 
information about the nature of the 
physical processes that produce the signals. 
For example, for either discrete or continuous
signals, that are functions of time, one may 
select a threshold value, record the time when the signals exceeds this threshold, wait some time during 
which the signal falls below this value, 
and then record the next time at which the 
signal exceeds the same threshold. 
This interval, the waiting time, is itself another random variable. 
The original random variable can be characterized 
by a {\it probability distribution} function, for example
a Gaussian distribution if the central limit theorem applies, 
or a non-Gaussian distribution with ``fat tails'' if extreme values have enhanced probability.
The distribution of waiting times is a distinct distribution, 
independent of the distribution of the original variable, and it has independent significance. 
If the successive waiting times are independent
and uncorrelated, the underlying processes is Poissonian, and the distribution of waiting times is expected to be exponential.
On the other hand if the waiting times have a 
``memory'' and are correlated, they
will be distributed according to a power law. 
Examination of waiting times to make this distinction has been a useful tool in the study of processes 
in geophysics~\citep{Lepreti2001ApJL}, 
space physics~\citep{Carbone2006PRL}, economics~\citep{GRECO2008PA}, and laboratory materials~\citep{Ferjani2008PRE}, to name a few.

We should note that it is not uncommon for signals to be correlated for small time separations (or lengths) up to a certain scale, and for larger scales to become uncorrelated.
In this case waiting time distribution would make a transition from power-law form for smaller separations (up to a correlation scale) and 
then transition to an exponential form. This appears to be the case for magnetic discontinuities in the solar wind measured by the PVI method at 1 au \citep{GrecoEA08}.
In particular, for waiting times corresponding 
to spatial scales of about 10$^6$ km or smaller,
the waiting times show a clear power-law distribution~\citep{GrecoEA09-ace}. 
For larger scales, it becomes exponential and therefore uncorrelated~\citep{GrecoEA09-wait}. 

Here, we will examine waiting time 
statistics for SEP 
data measured by the \isois/EPI-Lo instrument in the first PSP encounter. The purpose is to understand whether 
in this region, closer to the sun than previously explored, 
the occurrence of particle counts is random and uncorrelated, or
if the counts are clustered and correlated. 
Then, in later sections. we examine if apparent clustering is associated with the occurrence of magnetic discontinuities measured using the PVI method. Conclusions regarding 
these issues are likely to provide information valuable to discovering details of the acceleration and transport mechanisms responsible for observed SEP measurements. 

As a first analysis, we examine the 
distribution of count rates for
a particular channel of EPI-Lo data
during the first PSP solar encounter, which lasted from 31 October 2018 to 12 November 2018.
This is illustrated in Figure \ref{fig:hist-lo}.
One observes that low count rates are much more common than high rates, as expected for SEP data. 
This episodic 
property of energetic particle 
data makes analysis of statistical correlations between 
SEP counts and 
any ambient property such as magnetic field, 
particularly challenging.
In fact for very low count rates, there may be some 
question as to whether the 
signal is physically significant, or alternatively, 
one may be observing a noise signal, due to spurious fluctuations in electronics, for example. 

A definitive judgement as to whether
particular low-count signals are of significance
is difficult or even impossible.
However, a reasonable judgement may be made based
on the statistics of a particular data record:
If the signal exhibits correlations or ``clustering'',
then one might expect that its origin is systematic 
and likely of physical nature.
On the other hand, if the signal is consistent with uncorrelated events or Poisson noise, then one may suspect it is due to random noise or some other memory-less process.
In the latter case, it might not contain physical content, or, at least, 
it demonstrates that the physical processes at work are unrelated.
At low count rates Poisson signals would likely indicate the former
--- a lack of physical content, while at high count rate Poisson noise indicates that the physical processes measured are independent. 
On the other hand, non-Poissonian correlations are 
very likely to indicate physical correlations. 
As discussed above, an exponential distribution of waiting times between a signal exceeding a given threshold is associated with a Poisson signal, while a power-law distribution of waiting times suggests physical non-Poissonian correlations \citep{GrecoEA08}.

To examine \isois SEP data for Poissonianity versus clustering, we carry out several related tests of waiting times. 
First, for all EPI-Lo data in the selected channel during the first encounter, we compute the waiting time distribution between any two detected nonzero count-rates. Here, the episodic nature of energetic particle detection is a crucial element. Figure \ref{fig:wt0}
illustrates the result. A powerlaw fitting is obtained with Pearson's $r$ coefficient~\citep{PressEA} $r^2 = 0.95$. At the same time, an exponential fitting (not reported here) results in worse quality of fitting with $r^2=0.41$. It is apparent that the 
waiting time distribution is 
consistent with a power law, indicating 
clustering. 

As a next step, we examine waiting times  of \isois data by selecting the signals based on a threshold value of the count-rate. For low count-rate signals, we record the waiting times between signals which are less than $0.02\,\mathrm{s}^{-1}$. For high count-rate, we select the signals with greater than $0.1\,\mathrm{s}^{-1}$ count-rate. For both 
low count rate and high count rate signals, these conditional 
waiting time distributions are shown in Figure \ref{fig:wt-comb}. 
For low count rates, $<0.02\,\mathrm{s}^{-1}$,
we see in the left panel that the waiting time distribution is clearly exponential, supporting the conclusion that one-event level data is 
Poisson noise. The Pearson's $r$ coefficient is $r^2 = 0.99$ for exponential fit and $r^2 = 0.71$ for a powerlaw fit (not shown here), implying better agreement with the exponential fitting shown here. However, if the waiting times are computed for 
higher count rates $>0.1\,\mathrm{s}^{-1}$ only, one sees that 
a power-law  distribution is recovered. Here, although the number of points to fit is few $(4)$, the powerlaw fit yields a value of $r^2 = 0.97$, while an exponential fit (not shown) performs worse with $r^2 = 0.76$. Similar results are obtained for second solar encounter as well (not shown here).

\section{Energetic Particles and Coherent Structures} \label{sec:enc}
\begin{figure}
	\begin{center}
		\includegraphics[width=\linewidth]{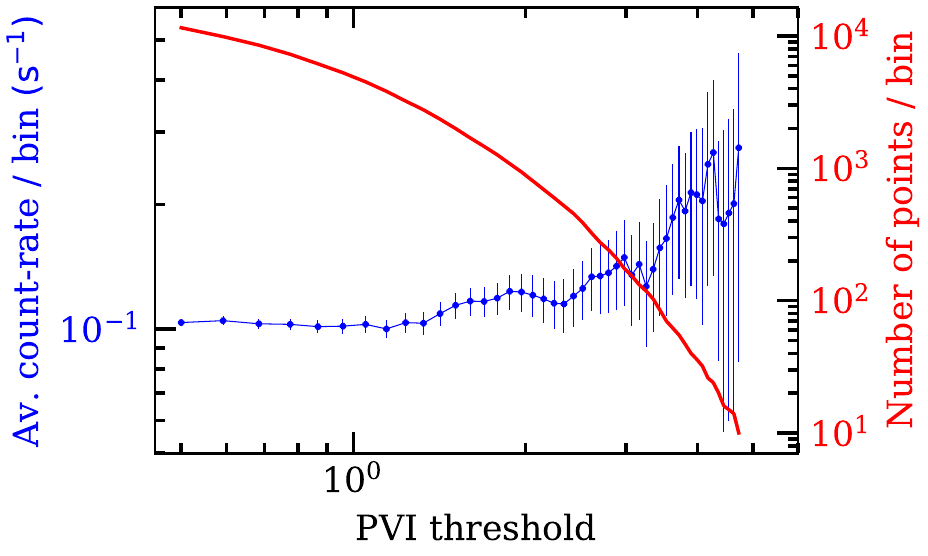}
		\caption{Average energetic particle count-rate plotted against PVI threshold for the first two solar encounters. A statistically significant increase in average count rates at higher PVI is noted. 		\label{fig:thres_enc}}
	\end{center}
\end{figure}
\begin{figure*}
	    \includegraphics[width=0.45\linewidth]{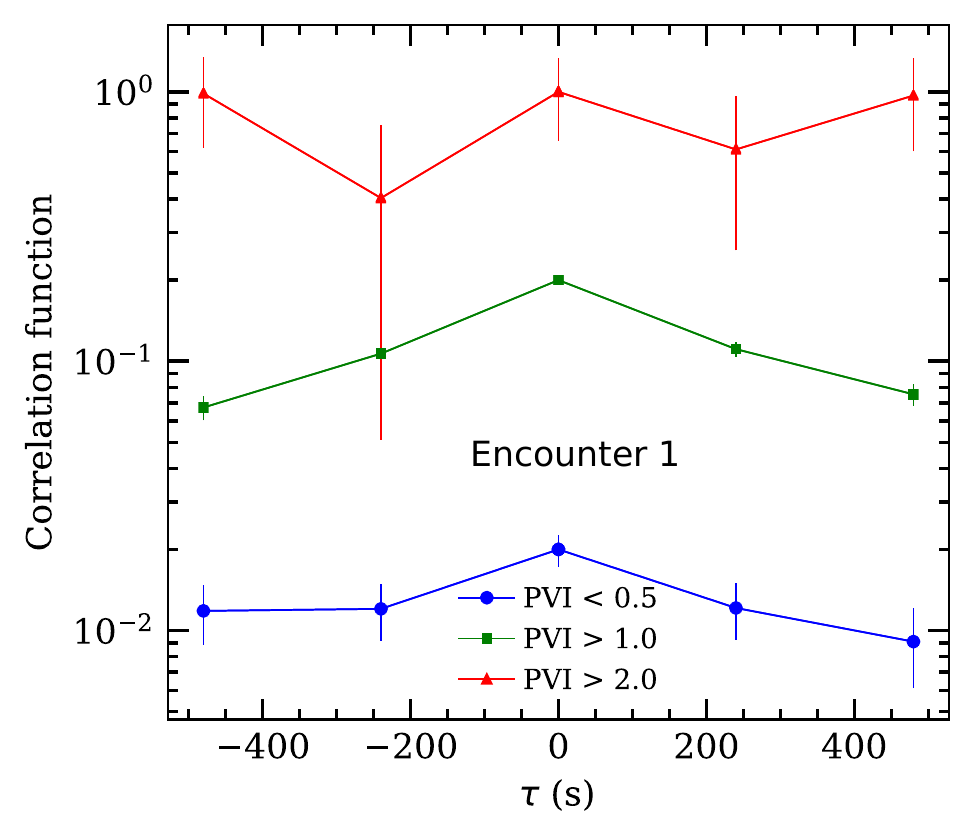}
		\includegraphics[width=0.45\linewidth]{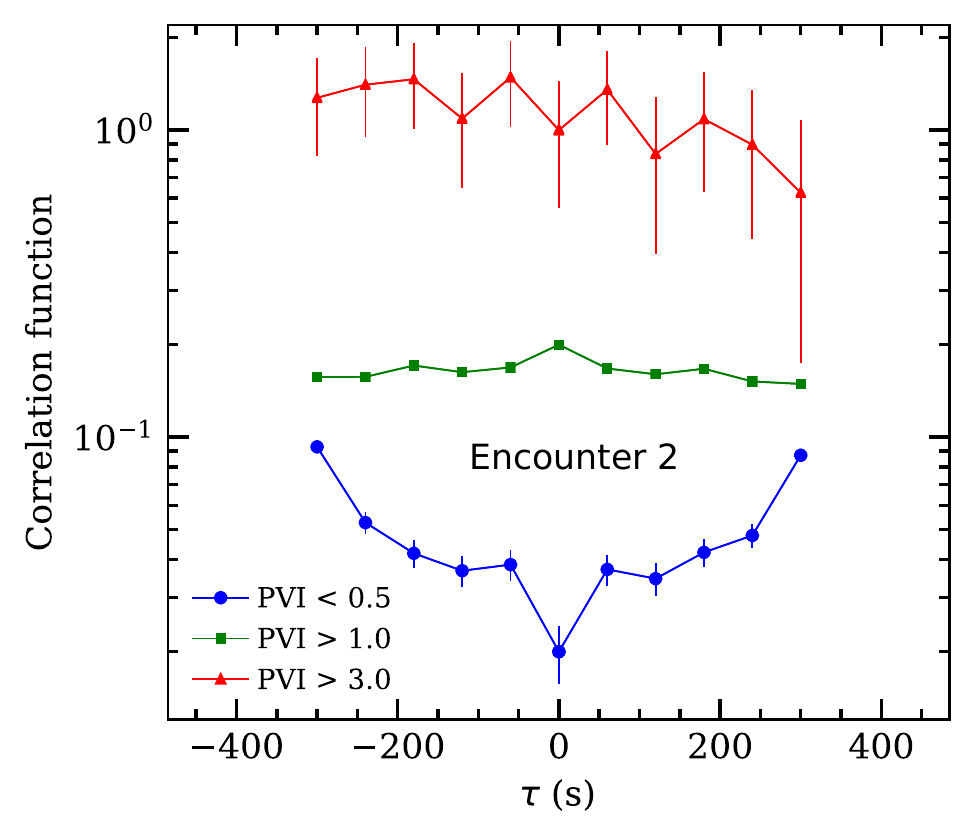}
		\caption{Time-lagged correlation of average energetic particle count-rate and PVI for different PVI threshold values for the first two encounters (left: encounter 1, right: encounter 2). The correlation functions have been shifted in the vertical direction for clarity. Note that the measurement cadence were 4 min and 1 min during the first and second encounter passes, respectively.}
		\label{fig:cond-stat}
\end{figure*}
A main challenge of statistical analysis of suprathermal-particle data is that these particles are already very rare. Therefore, usually long durations of data collection are 
required to deduce any statistical trend in the investigations. The first two solar encounters by PSP produce 
reasonably well 
populated SEP measurements to 
carry out some statistical correlations.

A turbulent system naturally generates \textit{patchy} or intermittent fluctuations and concentration of gradients of the 
primitive variables~\citep{SreenivasanAntonia97,Matthaeus2015PTRSA}. A practical technique for identifying these coherent structures is the method of \emph{Partial Variance Increment} (PVI)~\citep{GrecoEA08, GrecoEA09,Greco2017SSR}. This technique uses magnetic field data to identify small-scale structures such as current concentration. For single-spacecraft measurements, this method involves calculation of temporal increment of the magnetic field $|\Delta \mathbf{B}_{\tau} (t)| = |\mathbf{B} (t+\tau) - \mathbf{B} (t)|$. From that, the normalized partial variance of increments index (PVI index) at a lag of $\tau$ is given by
\begin{eqnarray}
\mathrm{PVI} (t) = \sqrt{\frac{|\Delta \mathbf{B}_{\tau} (t)|^2}{\langle |\Delta \mathbf{B}_{\tau} (t)|^2  \rangle}} \label{eq:pvi},
\end{eqnarray}
where, $\langle \cdots \rangle$ denotes a time average  over a suitably large trailing sample,
computed along the time series. Using the Taylor hypothesis~\citep{Taylor1938PRSLA}, then, one can convert temporal scales to length scales. 
At a heuristic level, one may think of PVI as a measurement
of the roughness of the magnetic field. Roughness will be greater in region where the fluctuations are larger, but PVI is most sensitive to relative roughness, in comparison to the regional values.
The increment of a turbulent field has long been of central importance in turbulence research, with particular importance having been
given to moments of the increment, the so-called structure functions \citep{MoninYaglom}. The square of PVI, as defined in equation~\ref{eq:pvi}, is related to the second-order structure function, but PVI is distinct in that it is a pointwise, rather than an averaged quantity. The PVI method
is one amongst several that have been developed
for identifying discontinuities in turbulent flows,
such as the Tsurutani-Smith method~\citep{Tsurutani1979JGR}, wavelet-based Local-Intermittency
Measure \citep{Veltri1999AIP,Farge2001PRL}, and the Phase Coherence Index \citep{Hada2003SSR}. \cite{Greco2017SSR} discuss a comparison of some of these methods with the PVI technique. We use the PVI method here for its simplicity and its direct relationship to increment statistics. {Here, we are interested in fluctuations near the inertial range of scales associated with the turbulence power spectrum~\cite{Kolmogorov1941c, Matthaeus1982aJGR}. The dynamics at these scales are governed by local-in-scale, non-linear processes and kinetic effects do not become dominant here. In this work, we calculate the PVI values for 2 min ($\sim\,$42,000 km $\sim\,$2800 $d_{\mathrm{i}}$) lag  which is well within
 the inertial range, for the first two encounters.} 
The averaging interval is 4 hours. Here, $d_{\mathrm{i}}$ is the ion-inertial length, defined as  $d_{\mathrm{i}} = c / \omega_{\mathrm{p i}} = \sqrt{m_{\mathrm{i}} \epsilon_{0} c^{2} / n_{\mathrm{i}} e^{2}}$, where $c$ is the speed of light in vacuum, $\omega_{\mathrm{p i}}$ is the proton plasma frequency, $m_{\mathrm{i}}$ is the proton mass, $\epsilon_{0}$ is the vacuum permittivity, $n_{\mathrm{i}}$ is the number density of protons, and $e$ is the proton charge. 

To study the association of energetic particle measurements with magnetic field structures, we calculate the average count-rate for a given PVI threshold value. The blue dots in figure~\ref{fig:thres_enc} plot the average energetic particle flux per PVI bin against PVI. Error bars are also shown as vertical lines. The uncertainty is estimated as $\sigma_{i} / \sqrt{m}$, where $\sigma_{i}$ is the standard deviation of the points in the $i^{\mathrm{th}}$ bin and $m$ is the number of points in that bin. The number of samples (right axis) for each PVI bin is shown for all data as a red, solid line. For PVI greater than 1 a rough, positive correlation with the count-rate can be observed,
with moderate statistical significance. 
Although, the error bars become large at higher PVI values $(> 3)$, on average energetic particle count rates and PVI threshold appear to be qualitatively correlated. This indicates that  there is a higher probability of finding high energetic particle count-rates near intense coherent structures.
A similar result was found at 1 au using ACE data~\citep{TesseinEA16}.

To further test the proximity of energetic particles with rough magnetic field structures, we calculate a time-lagged cross correlation between the two quantities. The normalized, cross correlation between two quantities, e.g., the EP count-rate $(C(t))$ and magnetic-field PVI, is defined as $R(\tau) / R(0)$, where,
\begin{eqnarray}
R(\tau) = \langle C(t)\,\mathrm{PVI}(t+\tau) + C(t+\tau)\,\mathrm{PVI}(t) \rangle \label{eq:corr}.
\end{eqnarray}
Here, $\langle \cdots \rangle$ denotes time average over the time series. Figure~\ref{fig:cond-stat} shows this cross correlation 
computed from subsets of the \isois data during the first two encounters 
for several PVI thresholds. The left panel shows the lagged correlation
for the first encounter, and the right panel shows the 
same quantity computed from the second encounter data. 
The different curves have been vertically shifted from the original value of unity at zero lag, for 
ease of visualization. The error bars are shown as vertical lines, when they are bigger than the plotting symbols. We only plot those points for which the relative errors are smaller than unity.

For the first two encounters, 
the correlation times are close to $\tau_{\mathrm{corr}} \sim 500\,\mathrm{s}$~\citep{Chhiber:pvi_WT} (in this volume). To avoid sampling the large-scale inhomogeneities, we calculate the cross correlations for only a fraction of the average correlation time $\tau < 500\,\mathrm{s}$.

\begin{figure}
	\begin{center}
		\includegraphics[width=\linewidth]{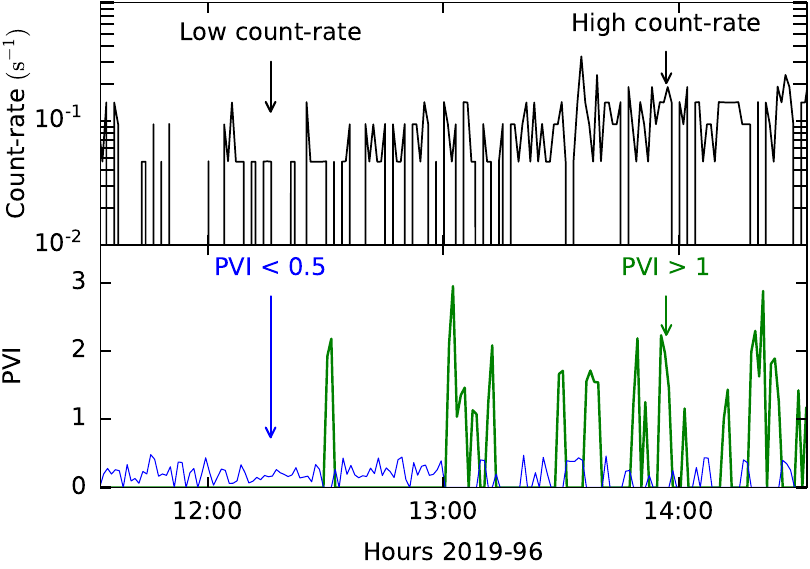}
		\caption{An example of observations of low (or high) energetic-particle count rate occurring near low (or high) PVI events, from encounter 2. The PVI $< 0.5$ values are shown in thin, blue line, and the PVI $> 1$ values are shown in thick, green line.}
		\label{fig:enc2_ex}
	\end{center}
\end{figure}

For the cases with PVI $>1$, for both encounters,
correlation peaks near zero lag, meaning that two quantities are correlated and changing together in time. 
However, an interesting trend can be observed from the PVI $<0.5$ curve for encounter 2. In this case, the correlation is slightly suppressed near zero lag and then increases far from the mid-point, in the direction of negative lags
(earlier times).
This indicates that very low SEP counts are 
found in regions of very smooth magnetic fields. 
Possible explanations
 for these correlations are discussed briefly in the Discussion section. A sample case for this type of correlation between energetic-particle data and PVI level is shown in figure~\ref{fig:enc2_ex}. The interval from encounter 2, shows that in the first part of interval, when the count rates are somewhat low, the PVI values are also small, on average. In the later part, both count rate and PVI values are, most of the time, high.

\begin{figure}
	\begin{center}
		\includegraphics[width=\linewidth]{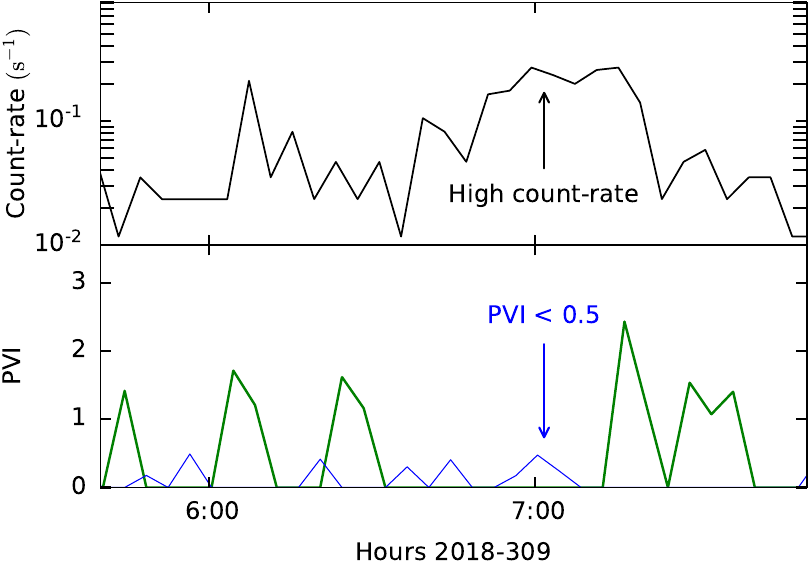}
		\caption{An example of observations of high energetic-particle count rate occurring near low PVI events, from encounter 1. Plotting conventions are same as figure~\ref{fig:enc2_ex}.}
		\label{fig:enc1_ex}
	\end{center}
\end{figure}

Conversely, in regions distant from a known smooth 
magnetic field location, one finds higher EP counts. 
On the other hand, the correlation for very low PVI in encounter 
1 actually {\it peaks} at zero lag. This implies that 
an enhanced 
energetic particle count rate may be  
found near even relatively weak magnetic field gradients. Figure~\ref{fig:enc1_ex} exhibits an example of an event when the count rate is high but the PVI is smaller than 0.5. One may note however that in this event the region of higher particle counts is approximately bounded on each side by elevated PVI (although still relatively low values).
This is reminiscent of the suggestion made by \citet{TesseinEA16} that PVI events at edges of particle enhancements may sometimes be a signature of confinement (see also \citet{Seripienlert2010ApJ}.)

Another contrasting feature from figure~\ref{fig:cond-stat} is that 
for encounter two, 
the PVI $>3$ curve is somewhat skewed with more average energetic particles before the PVI events (negative time lag) and fewer counts after (positive time lag). This trend can be qualitatively interpreted in the following way: Let us assume that, except for shocks, sharp magnetic discontinuities (high PVI) lie near the flux tube boundaries. Close to the Sun, where these structures are being accelerated rapidly, there may be a pile up of energetic particles just at the front edge of the boundaries while the density of energetic particles inside the flux tube would be low comparatively. 
This is an intriguing possibility, {though at 
present the results are qualitative, so a firm conclusion cannot be reached at this stage.}

These interesting features are only prominent for the second encounter analysis, presumably due to larger sample size. The first encounter results show peaked correlation at zero lag for all three PVI threshold values that are shown. Nevertheless, the rate of the decrease from zero lag, along both directions, become steeper for higher values of PVI threshold. 
We refrain from interpreting any deep conclusions from these results at this stage.


\section{A Dispersive Event} \label{sec:event}
\begin{figure}
	\begin{center}
		\includegraphics[width=\linewidth]{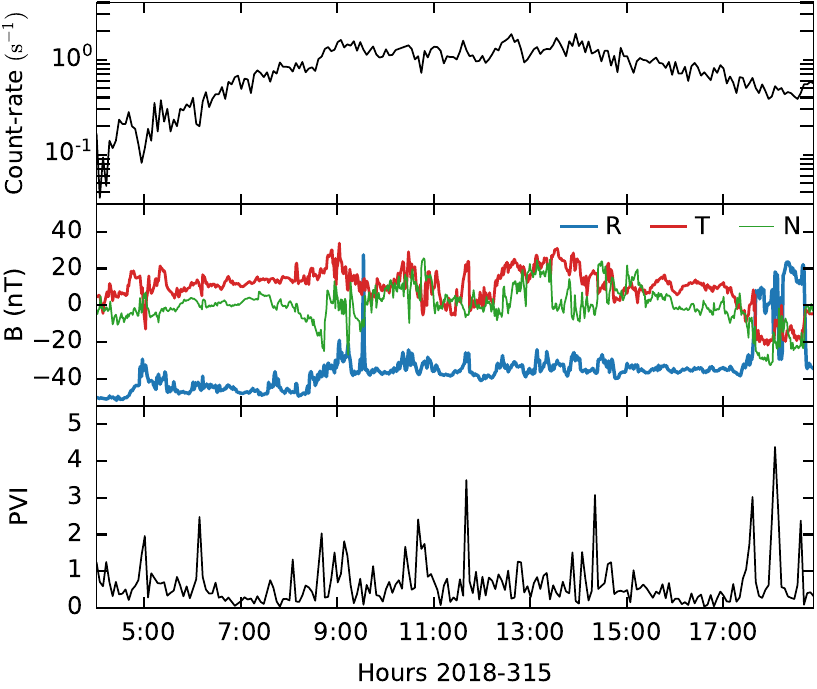}
		\caption{Time series plot of (top) EPI-Lo proton count rates, (middle) the magnetic-field components, and (bottom) the magnetic-field PVI.}
		\label{fig:overview_315}
	\end{center}
\end{figure}
A dispersive event, thought to be associated with a weak CME, was observed near the end of first encounter on 2018-315 from 04:00 to 19:00 UTC, approximately (see \cite{McComas:sub}, Giacalone et al. in this special issue, also see Korreck et al., Nieves-Chinchilla et al., Rouillard et al. in this volume). Figure~\ref{fig:overview_315} plots the time series of the relevant quantities for this time interval.  In this section, we repeat the analyses in the previous section for this particular event separately. 
Statistical correlations 
derived from data when the source is known to be relatively 
nearby,
such as this event, presents an interesting point of contrast
to the general statistics
derived from 
the entire encounter presented 
in the previous section. 

\begin{figure}
	\begin{center}
		\includegraphics[width=\linewidth]{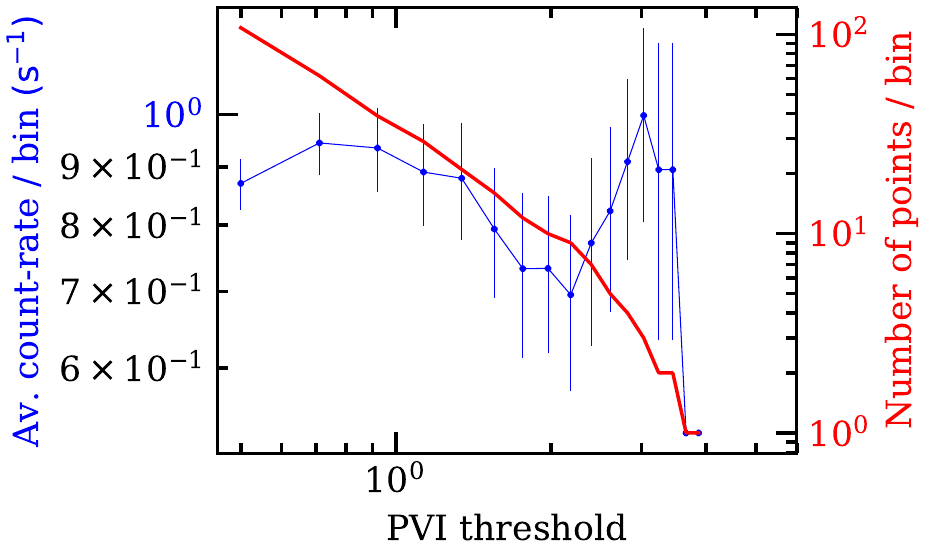}
		\caption{Average energetic particle count-rate plotted against PVI
threshold for the dispersive event on 2018-315.}
		\label{fig:315-thres}
	\end{center}
\end{figure}

Figure~\ref{fig:315-thres} employs the data from the dispersive event, and plots the conditionally averaged EP count-rate for various selected PVI thresholds, similar to figure~\ref{fig:thres_enc}.
Even though the calculations 
now involve only this limited data interval,
a positive correlation between average EP populations and coherent structures is again observed. 
While this result appears to be statistically significant, the 
regional (time-lagged) 
correlation results, derived from the dispersive event is rather weak and might be inconclusive;  therefore, it is discussed in a brief Appendix.

\section{Discussion and Conclusions}
In this paper, by 
performing a study of joint statistics of \isois/EPI-Lo samples, in conjunction with FIELDS/MAG measurements, we have presented first results on the association
of EP and magnetic field structures in the inner heliosphere, 
most likely close to the Alfv\'en surface. 

In a first step,
we used a waiting time analysis for different count-rate threshold, 
and reached a conclusion about the clustering or randomness of the samples.
Specifically, we find that 
all but the lowest count rate \isois data exhibit 
waiting times consistent with clustering, 
supporting the interpretation that physical correlations are detected. 

Several important extensions
to this simple exercise remain to be performed. Instead of categorizing the data by simply count-rate, one may undertake 
a more advanced survey, based on simultaneous EPI-Lo and EPI-Hi measurements. Due to several external factors, e.g., light contamination, temperature, cosmic rays, etc., some weak count-rates may be based on real activity, and may not just be 
due to background noise. A waiting time study, similar to the one presented here, might be able to differentiate these kind of signals.

Another task might be 
to compare the results for different energy ranges of the instrument. It is expected that, the method would be able to distinguish between counts in the very high energy range $( > 1\,\mathrm{MeV})$, which are presumably dominated by background noise, and below $ 200\,\mathrm{keV}$, which are expected to have foreground contamination. 

The second stage of the analysis 
examined the conditionally 
averaged energetic particle count rates, based
on a simplistic measure of the 
local intermittency, or roughness of the magnetic field.
The method employed a Partial Variance of Increment analysis, with the purpose of 
selecting structures with strong magnetic gradients, indicating the possible presence of coherent structures. The results for the first two PSP encounters appear to be consistent with the conclusion that solar energetic particles are likely to be correlated with coherent magnetic structures. This suggests the possibility that energetic particles might be concentrated near magnetic flux tube boundaries, although other interpretations are also possible \citep[see e.g.,][]{Kittinaradorn_2009,Seripienlert2010ApJ,Tooprakai2016ApJ,Malandraki2019ApJ}. This effect has also been reported at 1 au and may be associated either with transport effects, or even local acceleration processes~\citep{TesseinEA16, Khabarova2015ApJ}. An analysis focusing on a single day during which a dispersive SEP event occurred, suggests similar conclusions. However, to draw firmer conclusions we must await data from future PSP orbits. Other possible explanations may emerge.
For example, this inverse correlation for PVI $< 0.5$ might be an indication 
that pressure induced locally by energetic particles smooths out the field. 
The fact that this occurs prominently in the second solar encounter, may be an indication
 of increased solar activity during this encounter.
 Another possible interpretation is that the low PVI regions have low-beta structures, 
so the EP counts are weak in these regions. 
These ideas 
require further testing from SWEAP data, when available in the future. 
As further data from these type of events is 
accumulated by Parker Solar Probe and future missions like Solar Orbiter, and as regions even closer to the corona will be explored, 
it is likely 
that the crucial questions 
regarding the relative role of 
transport effects and sources of energization 
will be further clarified. 

\section*{Acknowledgments}
Parker Solar Probe was designed, built, and is now operated by the Johns Hopkins Applied Physics Laboratory as part of NASA’s Living with a Star (LWS) program (contract NNN06AA01C). Support from the LWS management and technical team has played a critical role in the success of the Parker Solar Probe mission. We are deeply indebted to everyone who helped make the PSP mission
possible. In particular, we thank all of the outstanding scientists, engineers, technicians, and
administrative support people across all of the \isois institutions that produced and supported the
\isois instrument suite and support its operations and the scientific analysis of its data. 
We also thank the FIELDS and SWEAP teams for 
cooperation. The \isois data and visualization tools are available to the community at: \href{https://spacephysics.princeton.edu/missions-instruments/isois}{https://spacephysics.princeton.edu/missions-instruments/isois}; data are also available via the NASA Space Physics Data Facility (\href{https://spdf.gsfc.nasa.gov/}{https://spdf.gsfc.nasa.gov/}). This research was partially supported by the Parker Solar Probe Plus project through Princeton/\isois 
subcontract SUB0000165, and in part by NSF-SHINE AGS-1460130 and grant RTA6280002 from Thailand Science Research and Innovation. S.D.B. acknowledges the support of the Leverhulme Trust Visiting Professorship program.


\section*{Appendix}
\begin{figure}
	\begin{center}
		\includegraphics[width=0.8\linewidth]{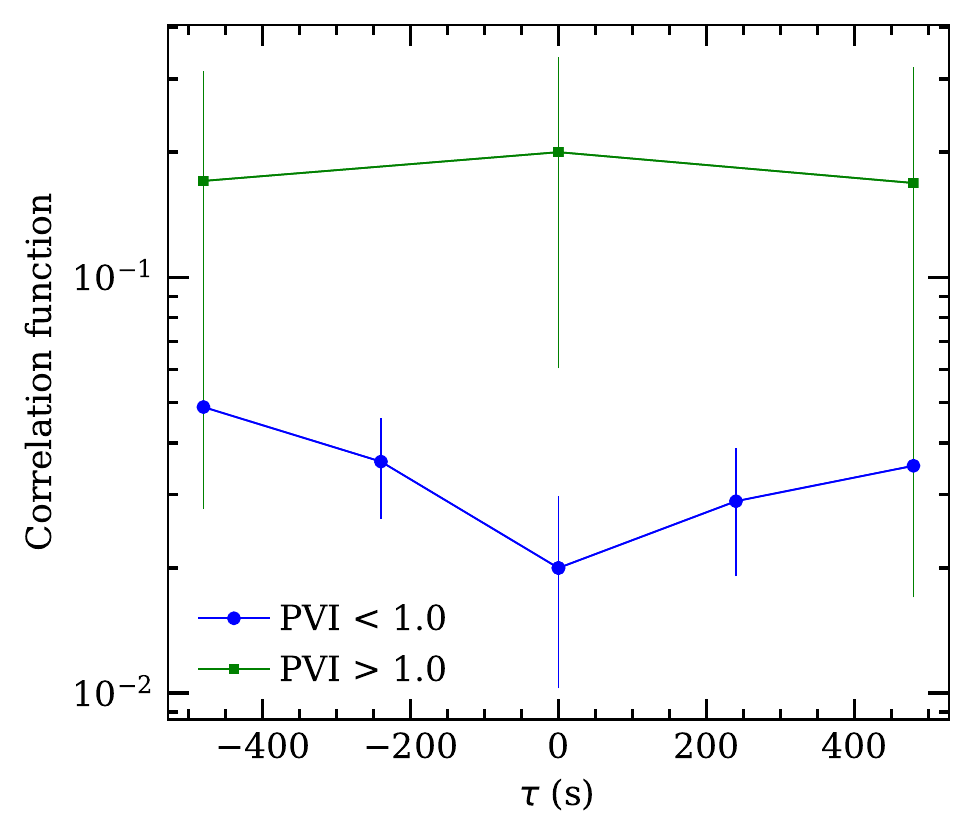}
		\caption{Time-lagged correlation of average energetic particle count-rate and PVI for different PVI threshold values for the dispersive event on 2018-315.}
		\label{fig:corr-315}
	\end{center}
\end{figure}

Here, we repeat a similar calculation like figure~\ref{fig:cond-stat}, but only for the dispersive event on 2018-315. The results in figure~\ref{fig:corr-315} show similar feature as those in figure~\ref{fig:cond-stat}, computed from longer data sets, encompassing the first two encounters. However, due to low sample size, the prominence is weak and the conclusions are qualitative, at best.


\begin{thebibliography}{}
	\expandafter\ifx\csname natexlab\endcsname\relax\def\natexlab#1{#1}\fi
	

	\bibitem[{Amato \& Blasi(2018)}]{Amato2018ASR}

	Amato, E., \& Blasi, P. 2018, Advances in Space Research, 62, 2731 , origins of
	Cosmic Rays
	
	\bibitem[{Ambrosiano {et~al.}(1988)Ambrosiano, Matthaeus, Goldstein, \&
		Plante}]{AmbrosianoEA88}
	Ambrosiano, J., Matthaeus, W.~H., Goldstein, M.~L., \& Plante, D. 1988, Journal
	of Geophysical Research: Space Science, 93, 14\,383
	
	\bibitem[{Bale {et~al.}(2016)Bale, Goetz, Harvey, Turin, Bonnell,
		Dudok de Wit, Ergun, MacDowall, Pulupa, Andre, Bolton, Bougeret, Bowen,
		Burgess, Cattell, Chandran, Chaston, Chen, Choi, Connerney, Cranmer,
		Diaz-Aguado, Donakowski, Drake, Farrell, Fergeau, Fermin, Fischer, Fox,
		Glaser, Goldstein, Gordon, Hanson, Harris, Hayes, Hinze, Hollweg, Horbury,
		Howard, Hoxie, Jannet, Karlsson, Kasper, Kellogg, Kien, Klimchuk,
		Krasnoselskikh, Krucker, Lynch, Maksimovic, Malaspina, Marker, Martin,
		Martinez-Oliveros, McCauley, McComas, McDonald, Meyer-Vernet, Moncuquet,
		Monson, Mozer, Murphy, Odom, Oliverson, Olson, Parker, Pankow, Phan,
		Quataert, Quinn, Ruplin, Salem, Seitz, Sheppard, Siy, Stevens, Summers,
		Szabo, Timofeeva, Vaivads, Velli, Yehle, Werthimer, \& Wygant}]{Bale2016SSR}
	Bale, S.~D., Goetz, K., Harvey, P.~R., {et~al.} 2016, Space Science Reviews,
	204, 49
	         
    \bibitem[{Bale {et~al.}(2019)Bale, Badman, Bonnell}]{Bale:sub}
    Bale, S.~D., Badman, S.~T., Bonnell, J.~W., {et~al.} 2019, Nature,
    1476, doi:10.1038/s41586-019-1818-7         
	
	\bibitem[{Bieber {et~al.}(1994)Bieber, Matthaeus, Smith, Wanner, Kallenrode, \&
		Wibberenz}]{BieberEA94}
	Bieber, J.~W., Matthaeus, W.~H., Smith, C.~W., {et~al.} 1994, The Astrophysical
	Journal, 420, 294
	
	\bibitem[{Carbone {et~al.}(2006)Carbone, Sorriso-Valvo, Vecchio, Lepreti,
		Veltri, Harabaglia, \& Guerra}]{Carbone2006PRL}
	Carbone, V., Sorriso-Valvo, L., Vecchio, A., {et~al.} 2006, Phys. Rev. Lett.,
	96, 128501
	
	\bibitem[{{Chasapis} {et~al.}(2015){Chasapis}, {Retino}, {Sahraoui}, {Vaivads},
		{Khotyaintsev}, {Sundkvist}, {Greco}, {Sorriso-Valvo}, \&
		{Canu}}]{Chasapis2015ApJL}
	{Chasapis}, A., {Retino}, A., {Sahraoui}, F., {et~al.} 2015, Astrophys. J.
	Lett., 804, L1
	
	\bibitem[{{Chhiber} {et~al.}(2020){Chhiber}, Goldstein, Maruca, Chasapis,
		Matthaeus, Ruffolo, Bandyopadhyay, Parashar, Qudsi, Dudok~de Wit, Bale,
		Bonnell, Goetz, Harvey, , MacDowall, Malaspina, Pulupa, Kasper, Korreck,
		Case, Stevens, Whittlesey, Larson, Livi, Velli, \& Raouafi}]{Chhiber:pvi_WT}
	{Chhiber}, C., Goldstein, M.~L., Maruca, B.~A., {et~al.} 2020, in press.
	
	\bibitem[{{Dmitruk} {et~al.}(2004){Dmitruk}, {Matthaeus}, \&
		{Lanzerotti}}]{DmitrukEA04}
	{Dmitruk}, P., {Matthaeus}, W.~H., \& {Lanzerotti}, L.~J. 2004, \grl, 31, 21805
	
	\bibitem[{Farge {et~al.}(2001)Farge, Pellegrino, \& Schneider}]{Farge2001PRL}
	Farge, M., Pellegrino, G., \& Schneider, K. 2001, Phys. Rev. Lett., 87, 054501
	
	\bibitem[{Ferjani {et~al.}(2008)Ferjani, Sorriso-Valvo, De~Luca, Barna,
		De~Marco, \& Strangi}]{Ferjani2008PRE}
	Ferjani, S., Sorriso-Valvo, L., De~Luca, A., {et~al.} 2008, Phys. Rev. E, 78,
	011707
	
	\bibitem[{Fox {et~al.}(2016)Fox, Velli, Bale, Decker, Driesman, Howard, Kasper,
		Kinnison, Kusterer, Lario, Lockwood, McComas, Raouafi, \& Szabo}]{Fox2016SSR}
	Fox, N.~J., Velli, M.~C., Bale, S.~D., {et~al.} 2016, Space Science Reviews,
	204, 7
	
	\bibitem[{Greco {et~al.}(2008{\natexlab{a}})Greco, Chuychai, Matthaeus,
		Servidio, \& Dmitruk}]{GrecoEA08}
	Greco, A., Chuychai, P., Matthaeus, W.~H., Servidio, S., \& Dmitruk, P.
	2008{\natexlab{a}}, Geophysical Research Letters, 35,
	doi:10.1029/2008GL035454
	
	\bibitem[{Greco {et~al.}(2009)Greco, Matthaeus, Servidio, Chuychai, \&
		Dmitruk}]{GrecoEA09}
	Greco, A., Matthaeus, W.~H., Servidio, S., Chuychai, P., \& Dmitruk, P. 2009,
	The Astrophysical Journal, 691, L111
	
	\bibitem[{{Greco} {et~al.}(2009{\natexlab{a}}){Greco}, {Matthaeus}, {Servidio},
		{Chuychai}, \& {Dmitruk}}]{GrecoEA09-ace}
	{Greco}, A., {Matthaeus}, W.~H., {Servidio}, S., {Chuychai}, P., \& {Dmitruk},
	P. 2009{\natexlab{a}}, The Astrophysical Journal Letters, 691, L111
	
	\bibitem[{{Greco} {et~al.}(2009{\natexlab{b}}){Greco}, {Matthaeus}, {Servidio},
		\& {Dmitruk}}]{GrecoEA09-wait}
	{Greco}, A., {Matthaeus}, W.~H., {Servidio}, S., \& {Dmitruk}, P.
	2009{\natexlab{b}}, Phys. Rev. E, 80, 046401
	
	\bibitem[{Greco {et~al.}(2008{\natexlab{b}})Greco, Sorriso-Valvo, Carbone, \&
		Cidone}]{GRECO2008PA}
	Greco, A., Sorriso-Valvo, L., Carbone, V., \& Cidone, S. 2008{\natexlab{b}},
	Physica A: Statistical Mechanics and its Applications, 387, 4272
	
		\bibitem[{Greco {et~al.}(2017)Greco, {et al.}}]{Greco2017SSR}
	Greco. A., Matthaeus, W. H., Perri, S., Osman, K. T., Servidio, S., Wan, M., 
           \& Dmitrik, P. 2017, Space Science Reviews, 214, 1

	\bibitem[{Hada {et~al.}(2003)Hada, Koga, \& Yamamoto}]{Hada2003SSR}
	Hada, T., Koga, D., \& Yamamoto, E. 2003, Space Science Reviews, 107, 463
	
	\bibitem[{Jokipii(1966)}]{Jokipii66}
	Jokipii, J.~R. 1966, The Astrophyscial Journal, 146, 480
	
	\bibitem[{Khabarova {et~al.}(2015)Khabarova, Zank, Li, le~Roux, Webb, Dosch, \&
		Malandraki}]{Khabarova2015ApJ}
	Khabarova, O., Zank, G.~P., Li, G., {et~al.} 2015, The Astrophysical Journal,
	808, 181
	
	\bibitem[{Khabarova \& Zank(2017)}]{Khabarova2017ApJ}
	Khabarova, O.~V., \& Zank, G.~P. 2017, The Astrophysical Journal, 843, 4
	
	\bibitem[{Kittinaradorn {et~al.}(2009)Kittinaradorn, Ruffolo, \&
		Matthaeus}]{Kittinaradorn_2009}
	Kittinaradorn, R., Ruffolo, D., \& Matthaeus, W.~H. 2009, The Astrophysical
	Journal, 702, L138
	
	\bibitem[{Kolmogorov(1941)}]{Kolmogorov1941c}
	Kolmogorov, A.~N. 1941, C.R. Acad. Sci. U.R.S.S., 32, 16, [Reprinted in Proc.\
	R.\ Soc.\ London, Ser.\ A \textbf{434}, 15--17 (1991)]
	
	\bibitem[{Lepreti {et~al.}(2001)Lepreti, Carbone, \& Veltri}]{Lepreti2001ApJL}
	Lepreti, F., Carbone, V., \& Veltri, P. 2001, The Astrophysical Journal, 555,
	L133
	
	\bibitem[{Loureiro {et~al.}(2012)Loureiro, Samtaney, Schekochihin, \&
		Uzdensky}]{LoureiroEA12}
	Loureiro, N.~F., Samtaney, R., Schekochihin, A.~A., \& Uzdensky, D.~A. 2012,
	Physics of Plasmas, 19, 042303
	
	\bibitem[{Malandraki {et~al.}(2019)Malandraki, Khabarova, Bruno, Zank, Li,
		Jackson, Bisi, Greco, Pezzi, Matthaeus, Giannakopoulos, Servidio, Malova,
		Kislov, Effenberger, le~Roux, Chen, Hu, \& Engelbrecht}]{Malandraki2019ApJ}
	Malandraki, O., Khabarova, O., Bruno, R., {et~al.} 2019, The Astrophysical
	Journal, 881, 116
	
	\bibitem[{Matthaeus \& Goldstein(1982)}]{Matthaeus1982aJGR}
	Matthaeus, W.~H., \& Goldstein, M.~L. 1982, J. Geophys. Res., 87, 6011
	
	\bibitem[{Matthaeus {et~al.}(2015)Matthaeus, Wan, Servidio, Greco, Osman,
		Oughton, \& Dmitruk}]{Matthaeus2015PTRSA}
	Matthaeus, W.~H., Wan, M., Servidio, S., {et~al.} 2015, Philosophical
	Transactions of the Royal Society of London A: Mathematical, Physical and
	Engineering Sciences, 373, doi:10.1098/rsta.2014.0154
	
	\bibitem[{McComas {et~al.}(2019)McComas, Christian, Cohen}]{McComas:sub}
	McComas, D.~J., Christian, E.~R., Cohen, C.~M.~S., {et~al.} 2019, Nature,
	1476-4687, doi:10.1038/s41586-019-1811-1
	
	\bibitem[{McComas {et~al.}(2016)McComas, Alexander, Angold, Bale, Beebe,
		Birdwell, Boyle, Burgum, Burnham, Christian, Cook, Cooper, Cummings, Davis,
		Desai, Dickinson, Dirks, Do, Fox, Giacalone, Gold, Gurnee, Hayes, Hill,
		Kasper, Kecman, Klemic, Krimigis, Labrador, Layman, Leske, Livi, Matthaeus,
		McNutt, Mewaldt, Mitchell, Nelson, Parker, Rankin, Roelof, Schwadron,
		Seifert, Shuman, Stokes, Stone, Vandegriff, Velli, von Rosenvinge, Weidner,
		Wiedenbeck, \& Wilson}]{McComas2016SSR}
	McComas, D.~J., Alexander, N., Angold, N., {et~al.} 2016, Space Science
	Reviews, 204, 187
	
	\bibitem[{Monin \& Yaglom(1971, 1975)}]{MoninYaglom}
	Monin, A.~S., \& Yaglom, A.~M. 1971, 1975, Statistical Fluid Mechanics, Vols 1
	and 2 (Cambridge, Mass.: MIT Press)
	
	\bibitem[{Osman {et~al.}(2014)Osman, Matthaeus, Gosling, Greco, Servidio, Hnat,
		Chapman, \& Phan}]{Osman2014PRL}
	Osman, K.~T., Matthaeus, W.~H., Gosling, J.~T., {et~al.} 2014, Phys. Rev.
	Lett., 112, 215002
	
	\bibitem[{Osman {et~al.}(2011)Osman, Matthaeus, Greco, \&
		Servidio}]{OsmanEA11-swHeat}
	Osman, K.~T., Matthaeus, W.~H., Greco, A., \& Servidio, S. 2011, The
	Astrophysical Journal, 727, L11
	
	\bibitem[{Press {et~al.}(1992)Press, Teukolsky, Vetterling, \&
		Flannery}]{PressEA}
	Press, W.~H., Teukolsky, S.~A., Vetterling, W.~T., \& Flannery, B.~P. 1992,
	Numerical Recipes: {The} Art of Scientific Computing (New York)
	
	\bibitem[{Reames(1999)}]{Reames1999SSR}
	Reames, D.~V. 1999, Space Science Reviews, 90, 413
	
	\bibitem[{{Seripienlert} {et~al.}(2010){Seripienlert}, {Ruffolo}, {Matthaeus},
		\& {Chuychai}}]{Seripienlert2010ApJ}
	{Seripienlert}, A., {Ruffolo}, D., {Matthaeus}, W.~H., \& {Chuychai}, P. 2010,
	The Astrophysical Journal, 711, 980
	
	\bibitem[{Servidio {et~al.}(2011)Servidio, Greco, Matthaeus, Osman, \&
		Dmitruk}]{Servidio2011JGR}
	Servidio, S., Greco, A., Matthaeus, W.~H., Osman, K.~T., \& Dmitruk, P. 2011,
	Journal of Geophysical Research: Space Physics, 116, doi:10.1029/2011JA016569
	
	\bibitem[{Sreenivasan \& Antonia(1997)}]{SreenivasanAntonia97}
	Sreenivasan, K.~R., \& Antonia, R.~A. 1997, 29, 435
	
	\bibitem[{Taylor(1938)}]{Taylor1938PRSLA}
	Taylor, G.~I. 1938, Proceedings of the Royal Society of London Series A, 164,
	476
	
	\bibitem[{Terasawa \& Scholer(1989)}]{Terasawa1989Science}
	Terasawa, T., \& Scholer, M. 1989, Science, 244, 1050
	
	\bibitem[{Tessein {et~al.}(2013)Tessein, Matthaeus, Wan, Osman, Ruffolo, \&
		Giacalone}]{Tessein2013ApJL}
	Tessein, J.~A., Matthaeus, W.~H., Wan, M., {et~al.} 2013, The Astrophysical
	Journal Letters, 776, L8
	
	\bibitem[{{Tessein} {et~al.}(2016){Tessein}, {Ruffolo}, {Matthaeus}, \&
		{Wan}}]{TesseinEA16}
	{Tessein}, J.~A., {Ruffolo}, D., {Matthaeus}, W.~H., \& {Wan}, M. 2016,
	Geophys. Res. Lett., 43, 3620
	
	\bibitem[{Tessein {et~al.}(2015)Tessein, Ruffolo, Matthaeus, Wan, Giacalone, \&
		Neugebauer}]{Tessein2015ApJ}
	Tessein, J.~A., Ruffolo, D., Matthaeus, W.~H., {et~al.} 2015, The Astrophysical
	Journal, 812, 68
	
	\bibitem[{Tooprakai {et~al.}(2016)Tooprakai, Seripienlert, Ruffolo, Chuychai,
		\& Matthaeus}]{Tooprakai2016ApJ}
	Tooprakai, P., Seripienlert, A., Ruffolo, D., Chuychai, P., \& Matthaeus, W.~H.
	2016, The Astrophysical Journal, 831, 195
	
	\bibitem[{Tsurutani \& Smith(1979)}]{Tsurutani1979JGR}
	Tsurutani, B.~T., \& Smith, E.~J. 1979, Journal of Geophysical Research: Space
	Physics, 84, 2773
	
	\bibitem[{Veltri \& Mangeney(1999)}]{Veltri1999AIP}
	Veltri, P., \& Mangeney, A. 1999, AIP Conference Proceedings, 471, 543
	
	\bibitem[{{Wan} {et~al.}(2013){Wan}, {Matthaeus}, {Servidio}, \&
		{Oughton}}]{WanEA13-xpoints}
	{Wan}, M., {Matthaeus}, W.~H., {Servidio}, S., \& {Oughton}, S. 2013, Physics
	of Plasmas, 20, 042307
	
	\bibitem[{Zank {et~al.}(2014)Zank, le~Roux, Webb, Dosch, \&
		Khabarova}]{Zank2014ApJ}
	Zank, G.~P., le~Roux, J.~A., Webb, G.~M., Dosch, A., \& Khabarova, O. 2014, The
	Astrophysical Journal, 797, 28
	
	\bibitem[{Zhdankin {et~al.}(2012)Zhdankin, Boldyrev, Mason, \&
		Perez}]{ZhdankinEA12}
	Zhdankin, V., Boldyrev, S., Mason, J., \& Perez, J.~C. 2012, Phys. Rev. Lett.,
	108, 175004
	
	\bibitem[{Zhdankin {et~al.}(2013)Zhdankin, Uzdensky, Perez, \&
		Boldyrev}]{ZhdankinEA13}
	Zhdankin, V., Uzdensky, D.~A., Perez, J.~C., \& Boldyrev, S. 2013, The
	Astrophysical Journal, 771, 124
	
\end{thebibliography}

\end{document}